\documentstyle[prd,eqsecnum,preprint,tighten,aps,amsfonts]{revtex}
\newcommand{\sump}{\sum_{n=0}^\infty \!{}^{'}}

\renewcommand{\Re}{\mbox{Re }}

\begin{document}
\draft
\title{Casimir effect for a perfectly  conducting wedge in terms of
local zeta function}
\author{V.V.~Nesterenko\thanks{Electronic address: nestr@thsun1.jinr.ru}}
\address{Bogoliubov Laboratory of Theoretical Physics,
Joint Institute for Nuclear Research, 141980 Dubna,  Russia}
\author{G.~Lambiase\thanks{Electronic address: lambiase@sa.infn.it}
and G. Scarpetta\thanks{Electronic address: scarpetta@sa.infn.it}}
\address{Dipartimento di Fisica ``E.R. Caianiello'',
 Universit\'a di Salerno, 84081 \\ Baronissi (SA), Italy,\\
 INFN, Gruppo Collegato di Salerno, Italy,\\
and International Institute for Advanced Scientific
Studies--Vietri sul Mare, Italy}
\date{February 14, 2002}
\maketitle
\begin{abstract}
    The vacuum energy density of electromagnetic field inside a
perfectly conducting wedge is calculated by making use of the local
zeta function technique.  This regularization completely eliminates
divergent expressions in the course of calculations and gives rise to
a finite expression for the energy density in question without any
subtractions.  Employment of the Hertz potentials for constructing
the general solution to the Maxwell equations results in a
considerable simplification of the calculations.  Transition to the
global zeta function is carried out by introducing a cutoff nearby the
cusp at the origin.
Proceeding from this the heat kernel coefficients are
calculated and the high temperature asymptotics of the Helmholtz free
energy and of the torque of the Casimir forces are found. The wedge
singularity gives rise to a specific high temperature behaviour $\sim
T^2$ of the quantities under consideration. The obtained results are
directly applicable to the free energy of a scalar massless field and
electromagnetic field on
the background of a cosmic string.
\end{abstract}


\section{Introduction}
The Casimir calculations are usually accomplished by two
methods: a global  approach, when from the very
beginning the total vacuum energy of quantum fields are calculated
for given boundary conditions, and a local consideration, when first the
space density of the vacuum energy is found. The both methods are
in agreement if the integration of the vacuum energy density gives
the total Casimir energy. As known, the first example where this
condition is not satisfied was a perfectly conducting wedge, i.e.,
two perfectly conducting plates that are crossed at given angle.
The calculation of the Casimir energy   for these boundary conditions
has the following
peculiarities. The spectrum of electromagnetic oscillations inside the wedge
is continues and it does not depend explicitly on the opening angle
of the wedge. The boundaries are flat therefore there
is no problem with additional divergences due to the boundary
curvature. However, on the cross line of the plates the smoothness
of the boundary is obviously violated due to the cusp. Till now a
correct definition of the total vacuum energy of the fields inside
the wedge is lacking. It has been calculated only the spatial
density of vacuum energy inside the wedge. It turns out that this
density possesses a non-integrable singularity $\sim r^{-4}$ in
cylindrical coordinates  $r, \theta, z$ with the $z$ axes along
the cross line of the plates. For the first time these
calculations have been done probably in Ref.\ \cite{LR} by making
use of a simple method of images and later in Ref.\ \cite{DC} by
applying the  Green's function method. In recent publications
\cite{BL,BP} in the framework of the Schwinger source theory it
was shown how to avoid some mathematical inconsistency taking
place in previous calculations \cite{DC,HK}.

The geometry of space analogous to that between the crossing
plates also encounters in other physical problems, in particular,
in the cosmic string theory (the space outside the string) and
under consideration of quantum fields at finite temperature
on the Rindler space-time.

The geometry of space in the background of a cosmic string is
determined by the solution of the Einstein equations with the
energy-momentum tensor of matter defined by the string. When the
string is located along the $z$ axes and possesses the linear mass
density $\mu $, then the space outside the string is isomorphic to
the manifold $ C_{2\pi - \phi}\times {\Bbb R}^1$, where ${\Bbb R}^1$
is an infinite line
along the $z$ axes and instead of a plane perpendicular to this
axes one has two dimensional cone $C_{2\pi-\phi}$  with the angle
deficiency $\phi =8 \pi G \mu$, $G$ being the gravitational
constant. The cone of an angle $\beta $ becomes a plane when $\beta =2\pi$.
Certainly  the boundary conditions for fields inside the
wedge $W_\alpha \times {\Bbb R}^1$ and on  the space $ C_\beta
\times{\Bbb R}^1 $
are in the general case different. For the cone $C_\beta $ the
periodic boundary conditions are natural. For fields inside the
wedge the boundary conditions are  determined by the relevant
physical origin of the fields and by the physical characteristics
of the boundary.

The manifold $C_\beta \times {\Bbb R} ^2$ arises under consideration of
fields on the Rindler space-time at finite temperature $\beta ^{-1}$.
In order to calculate the Helmholtz free energy and entropy  of fields
on this space the
local zeta function technique was proposed \cite{ZCV}. A
nontrivial problem encountered here is the construction of
required analytic continuation of this zeta function. Such a
continuation has been accomplished separately for lower and higher
eigenvalues of the Laplace operator in the problem at hand.

As far as we know the Casimir energy for a perfectly conducting
wedge at finite temperature was considered only in Ref.\ \cite{BD}
by the multiple scattering method. In that paper unusual
temperature dependence of the Casimir forces in this problem  was
noted. These forces, tending to reduce the angle $\alpha $ between the
plates, diminish  when taking into account the first temperature
correction.

The present article seeks to show that the most simple and
consistent (from the mathematical point of view) approach to the
calculation of the Casimir effect for a perfectly conducting wedge
is to use the local zeta function technique. In order to construct
such spectral function one should find not only the spectrum of
the regarding operator but also its eigenfunctions. When looking
for the solutions to the Maxwell equations  in confined regions
(waveguides  or resonators) it is convenient to use the Hertz
potentials or Hertz vectors \cite{DC,HdP}. This technique considerably
simplifies the  calculations, specifically, removing the problem
of choosing the gauge. It is this way that we shall follow when
constructing the local zeta function for electromagnetic field
inside a perfectly conducting wedge $W_\alpha \times {\Bbb R}^1$.

The layout of the paper is as follows. In Section II the Hertz
potentials are introduced and the solutions to Maxwell equations
inside a perfectly conducting wedge are constructed. These
potentials are expressed in terms of two scalar functions which
satisfy  Dirichlet and Neumann boundary conditions. In Section III
the local zeta function in the problem at hand is constructed. By
a simple change of angle variable this function reduces to the
analogous one for a massless scalar field on the manifold $C_\beta
\times {\Bbb R}^1$, when $\beta=2 \alpha $. Therefore the analytic
continuation developed in Ref.\ \cite{ZCV} is also applicable in our
case. The local zeta function constructed enables one to calculate the
vacuum energy density of the electromagnetic field inside a wedge without
subtraction of infinities.
In Section IV
transition to the
global zeta function is carried out by introducing a cutoff nearby the
cusp at the origin.
Proceeding from this the heat kernel coefficients are
calculated and the high temperature asymptotics of the Helmholtz free
energy and of the torque of the Casimir forces are found. The wedge
singularity gives rise to a specific high temperature behaviour $\sim
T^2$ of the quantities under consideration. The obtained results are
directly applicable to the free energy of a scalar massless field and
electromagnetic field on
the background of a cosmic string.
In Section V we summarize our results.
In the Appendix the local zeta functions are constructed for scalar massless
fields inside a wedge with Dirichlet and Neumann boundary conditions.

\section{Solution to Maxwell equations in terms of Hertz vectors}
Finding the solutions of Maxwell equations inside confined regions
proves to be a rather difficult problem. Mainly it is due to the
vector character of the electromagnetic field. A convenient
technique for this purpose is to use the electric (${\bf \Pi '}$)
and magnetic (${\bf \Pi ''}$) Hertz vectors
\cite{HdP,Stratton,FM}. The Lorentz gauge condition
$$
\text{div} {\bf A} +\frac{1}{c}\frac{\partial \, \varphi}{\partial
\,t}=0
$$
is satisfied identically because the Hertz vectors are connected
with the potentials ${\bf A}$ and $\varphi $ in the following way
$$
{\bf A} = \frac{1}{c}\frac{\partial\, {\bf \Pi}}{\partial\,
t},\quad \varphi =-\text{div}\,{\bf \Pi}{.}
$$
For boundary problems with cylindrical symmetry the electric
(${\bf E }$) and magnetic (${\bf H}$) fields are expressed in
terms of the Hertz vectors having only one nonzero component
\begin{equation}
\label{eq2-1} {\bf \Pi '}=(0,\,0,\,U), \quad {\bf \Pi}
''=(0,\,0,\,V){.}
\end{equation}
Here cylindrical coordinates ($r,\theta,z$) are used with $z$ axes
directed along the intersection line of the wedge planes. The
scalar functions $U$ and $V$ are the eigenfunctions of the Laplace
operator and meet, respectively, Dirichlet and Neumann conditions
on the boundary $\partial \Gamma $
\begin{eqnarray}
\Delta U=\omega ^2 U, \quad \left .U \right |_{\partial \Gamma }=0{,}
\label{eq2-2} \\
\Delta V=\omega ^2 V, \quad \left .\frac{\partial V}{\partial
n}\right |_{\partial \Gamma }=0{.} \label{eq2-3}
\end{eqnarray}
When considering the electromagnetic field inside a perfectly
conducting dihedral
of opening angle $\alpha$ (for simplicity, conducting
wedge $W_\alpha$)
the eigenvalues of the operator $(-\Delta)$ are
\begin{equation}
\label{eq2-4}
\omega ^2 \equiv\omega^2(k,\lambda)=k^2+\lambda ^2,\quad
-\infty <k<\infty, \quad 0\leq \lambda <\infty{,}
\end{equation}
with $k$ being the wave vector along the $z$ direction. In all the
functions we have dropped for simplicity the time-dependent factor
$e^{i\omega t}$.  The spectrum (\ref{eq2-4}) does not depend
explicitly on the opening angle $\alpha $ of the wedge. This point
proves to be a crucial one when attempting to construct a global zeta
function in this problem (see Section IV).

The Dirichlet boundary value problem (\ref{eq2-2}) has the following
normalized
eigenfunctions ($E$-modes)
\[
u_{\lambda n k}(r,\theta, z)=
\frac{e^{ikz}}{\sqrt{\pi \alpha}}
 J_{np}(\lambda r)\sin (n p\, \theta){,}
\]
\begin{equation}
\label{eq2-5} 0\leq \theta\leq \alpha, \quad p=\pi /\alpha, \quad
n=1,2,\ldots , \quad 0\leq\lambda < \infty\,{.}
\end{equation}
For the Neumann boundary value problem (\ref{eq2-3}) the normalized
eigenfunctions
($H$-modes) have the form
\[
v_{\lambda n k} (r, \theta,z)=\eta_{n0}
\frac{e^{ikz}}{\sqrt{\pi \alpha}}
J_{np}(\lambda r) \cos (np \theta){,}
\]
\begin{equation}
\label{eq2-6}
n=0,1,2,\ldots , \quad 0\leq \lambda < \infty, \quad \eta _{n0}=
\cases{\frac{{\displaystyle 1}}{{\displaystyle  \sqrt 2}},\quad
 n=0,\cr 1,\quad  n=1,2,\ldots \, .\cr}
 \end{equation}
The eigenfunctions (\ref{eq2-5}) and (\ref{eq2-6}) meet the
orthogonality condition
\begin{equation}
\label{eq2-7}
\int^\infty_{-\infty} dz \int^\infty_0r\,dr\int^{\alpha}_0d\theta\,
f^*_{\lambda 'n'k'}(r,\theta ,z)
f_{\lambda n k}(r,\theta,z)=\delta_{nn'}
\frac{1}{\lambda}\,\delta(\lambda-\lambda ')\,\delta(k-k')
\end{equation}
and the completeness condition
\begin{equation}
\label{eq2-8}
\sum_{n=0\,(1)}^{\infty}\int^\infty_{-\infty} dk\int _0^\infty
\lambda\,  d\lambda \,
f^*_{\lambda n k}(r',\theta' ,z')
f_{\lambda n k}(r,\theta,z)=\frac{1}{r}\,\delta (r-r')\,\delta(z-z')
\,\delta(\theta -\theta ')\,{.}
\end{equation}
Here $f_{\lambda n k}(r,\theta,z)$ are the eigenfunctions
(\ref{eq2-5}) or (\ref{eq2-6}). The relations (\ref{eq2-7}) and
(\ref{eq2-8}) can be easily verified if one takes into account the
formula \cite{SM}
\[
\int^\infty_0 \lambda \,d \lambda \, J_q(\lambda r)J_q(\lambda r')=
\frac{1}{r}\,\delta (r-r')\,{.}
\]

For given Hertzian vectors ${\bf \Pi}'$ and ${\bf \Pi}''$ the electric
and magnetic fields are constructed by the formulas
 \begin{eqnarray}
 {\bf E}&= &{\bbox \nabla}\times {\bbox \nabla} \times {\bf \Pi}'\,,
 \quad {\bf H}=-i\omega {\bbox \nabla}\times {\bf \Pi}' \qquad
 (E\text{-modes})\,,\ \nonumber \\
 \label{eq2-9}
 {\bf E}& =& i\omega {\bbox \nabla} \times {\bf \Pi}''\,,
 \qquad {\bf H}={\bbox \nabla}\times {\bbox \nabla}\times
 {\bf \Pi}'' \qquad
 (H\text{-modes})\,.
\end{eqnarray}
It has been proved \cite{Heyn} that the superposition of these modes
gives the general solution to the Maxwell equations in the problem
under consideration.

However in the zeta function formalism there is no need to recover 
the electric and magnetic fields trough these formulas in order to 
construct the energy density or the energy momentum tensor as one 
proceeds when the Green's function technique is used \cite{DC,BL,BP}. 
It is an essential merit of the approach considered.

\section{Local Zeta Function for Perfectly Conducting Wedge}
At the beginning we remind  a few general formulas that define a
local zeta function. Let we have a differential operator $ L$ with
well posed eigenvalue problem (here it is convenient to use the
Dirac bracket notation)
\begin{equation}\label{eq3-1}
L f_n(x)=\lambda_n f_n(x) \quad \mbox{or}\quad
   L\vert n>=\lambda_n \vert n>\,.
\end{equation}
By making use of the completeness of the vector set $\{\vert n>\}$,
one can represent (formally) a unity operator $I$ acting in the
linear space of the vectors $\vert n>$ as follows
\begin{equation}\label{eq3-2}
  I=\sum_n \vert n>< n\vert \,.
\end{equation}
In view of this, we have for the inverse operator $L^{-1}$
\begin{equation}\label{3.3}
L^{-1}=\sum_n \frac{\vert n><n\vert}{\lambda_n}\,.
\end{equation}
It can be easily checked with allowance for Eq. (\ref{eq3-1}). For
the $s$th power of the inverse operator $L^{-1}$ we can write
\begin{equation}\label{eq3-4}
L^{-s}=\sum_n \frac{\vert n>< n\vert}{\lambda_n^s}\,.
\end{equation}
The local zeta function $\zeta(s, x)$ of the operator $L$
is a diagonal matrix element of the operator~$L^{-s}$
\begin{equation}\label{eq3-5}
  \zeta(s, x)=\sum_n\frac{<x\vert n><n\vert x>}{\lambda_n^s}=
  \sum_n\lambda_n^{-s}f_n^*(x)f_n(x)\,.
\end{equation}
The global zeta function
\begin{equation}\label{eq3-5a}
  \zeta(s)=\sum_n\lambda_n^{-s}
\end{equation}
is obtained by integration of $\zeta(s, x)$ over the whole space
\begin{equation}\label{eq3-6}
  \zeta(s)=\int \zeta(s, x)\,dx\,.
\end{equation}
Thus the local zeta function $\zeta(s, x)$ can be interpreted as a
spatial density for the global zeta function $\zeta(s)$.

When calculating the vacuum energy of a relativistic quantum
field $E_0$ by making use of the zeta function regularization
\cite{Od,Ten}, we have
\begin{equation}\label{eq3-7}
  E_0=\frac{1}{2}\, \zeta\left(s=-\frac{1}{2}\right)\,,
\end{equation}
with the operator $L$ being the spatial part of the relevant
differential operator defining the field dynamics in the problem
at hand. In this case $\lambda_n=\omega_n^2$, where $\omega_n$
are the classical  characteristic frequencies of the field under
study. In view of Eqs. (\ref{eq3-6}) and (\ref{eq3-7}), the quantity
$(1/2)\zeta(-1/2, x)$ can be interpreted as the vacuum energy
density calculated in the framework of the zeta function
regularization.
Besides that, it is easy to show that the vacuum expectation value
of the  canonical energy-momentum tensor
of non-interacting scalar field  can be represented in the form
\begin{equation}
\label{eq3-8a}
<0|T_{00}(x)|0>=\frac{1}{2}\sum_{n}\lambda_n^{1/2}f_n^*(x)f_n(x)=
\frac{1}{2}\zeta \left(-\frac{1}{2},x
\right ){.}
\end{equation}
In this case the operator $L$ is simply $-\Delta $.

The main difficulty in obtaining the local zeta
function (\ref{eq3-5}), as well as the global one (\ref{eq3-5a}), is
an analytic continuation of these formula to the region $\Re
\,s<1$.

It is remarkable that for constructing the local zeta function
$\zeta (s, x)$ only the spectrum and the eigenfunctions of the
relevant operator are needed. The explicit  form of the regarding
physical fields (for example, electromagnetic fields) is not
involved in these formulas. Therefore we shall not use Eq.
(\ref{eq2-9}) determining the electric and magnetic fields.
Instead of that we proceed from the spectrum (\ref{eq2-4}) in the
problem at hand and the respective eigenfunctions (\ref{eq2-5})
and (\ref{eq2-6}).

Substituting Eqs.\ (\ref{eq2-4}), (\ref{eq2-5}) and (\ref{eq2-6})
into the definition of the local zeta function (\ref{eq3-5}) we
obtain
\begin{eqnarray}
 \zeta (s, r, \theta, z)&=&
\frac{1}{\pi\alpha}\int_{-\infty}^{\infty}\!\!\!dk\int_0^\infty
\!\!\!  \frac{\lambda
  d\lambda}{(k^2+\lambda^2)^s}\left\{\frac{1}{2} J_0^2(\lambda r)+
\sum_{n=1}^\infty J_{np}^2(\lambda
  r)[\sin^2(np\theta)+\cos^2(np\theta)]
\right \} \nonumber \\
&=&
\frac{1}{\pi\alpha}\int_{-\infty}^{\infty}\!\!\!dk\int_0^\infty
\!\!\!  \frac{\lambda
  d\lambda}{(k^2+\lambda^2)^s}
\left [ \frac{1}{2}J_0^2(\lambda r)
+\sum_{n=1}^\infty J_{np}^2(\lambda r)
\right ]{.}
  \label{eq3-9}
\end{eqnarray}
Thus the local zeta function in the problem at hand is
independent of the angle variable $\theta$ and, that is natural,
of $z$ due to the homogeneity of the system in the $z$-direction.
For simplicity, we shall drop these arguments in what follows.

By making use of the value of the integral
\begin{equation}\label{eq3-10}
  \int_{-\infty}^{\infty}\frac{dk}{(k^2+\lambda^2)^s}=\sqrt{\pi}\lambda^{1-2s}
  \frac{\Gamma(s-1/2)}{\Gamma(s)}\,, \quad
  \Re \,s>\frac{1}{2}\,,
\end{equation}
we cast Eq. (\ref{eq3-9}) into the form
\begin{equation}\label{eq3-11}
  \zeta(s, r)=
  \frac{1}{\sqrt{\pi}\,\alpha}\frac{\Gamma(s-1/2)}{\Gamma(s)}
\sump
 \int_0^\infty \lambda^{2-2s}J_{np}^2(\lambda r)d\lambda\,.
\end{equation}
The prime on the summation sign means that the $n=0$ term is
taken with half weight.

Now we have to use the following integration formula (see
equation 6.574.2 with $\mu=\nu$ in Ref. \cite{GR})
 \[
  \int_0^\infty t^{-\lambda}J_\nu^2(\alpha t)dt=\frac{\alpha^{\lambda-1}
\Gamma(\lambda)
  \Gamma\left(\displaystyle{\frac{2\nu-\lambda+1}{2}}\right)}
  {2^\lambda\Gamma^2\left(\displaystyle{\frac{\lambda+1}{2}}\right)
  \Gamma\left(\displaystyle{\frac{2\nu+\lambda+1}{2}}\right)}\,,
 \]
 \begin{equation}\label{eq3-12}
  \Re\, (2\nu+1)>\Re\,\lambda>0, \quad \alpha>0\,.
\end{equation}
Applying this formula we can accomplish the integration in each
term in Eq. (\ref{eq3-11}) for the respective values of the
variable $s$ restricted by the equalities
\begin{equation}\label{eq3-13}
  1<\Re\,s<np+\frac{3}{2}\,,\quad n=0, 1, 2,\ldots \,.
\end{equation}
In these regions the integral (\ref{eq3-12}) converges. Therefore
the objections against using this formula brought up in Ref.
\cite{BL} are obviously not applicable here. As a result, we
obtain
\begin{equation}\label{eq3-14}
\zeta(s, r)=\frac{1}{2\pi\alpha
r^{3-2s}}\frac{\Gamma(s-1)}{\Gamma(s)}\sump
\frac{\Gamma(np+3/2-s)}{\Gamma(np+s-1/2)}\,.
\end{equation}
When deriving this equation the formula for the doubling of the
gamma function argument \cite{GR}
\begin{equation}\label{eq3-15}
  \Gamma(2z)=(2\pi)^{-1/2}2^{2z-1/2}\Gamma(z)\Gamma\left(z+\frac{1}{2}\right)
\end{equation}
has been used.

By making use of the  asymptotics of the ratio of two gamma functions
\cite{Olver}, one can easily show (see below) that
the series (\ref{eq3-14}) converges in the region
\begin{equation}\label{eq3-16}
  \Re\, s>\frac{3}{2}\,,
\end{equation}
while the first term with $n=0$  in the sum (\ref{eq3-14}) is defined in the
region
\begin{equation}\label{eq3-17}
  1<\Re\, s<\frac{3}{2}\,.
\end{equation}
Thus in the problem under study, there is no representation for
the local zeta function (\ref{eq3-14}) determining it as an
analytic function of the complex variable $s$ in a finite region.
The first term in Eq. (\ref{eq3-14}) is defined in the region
(\ref{eq3-17}) but the rest of the sum is defined in the strip
\begin{equation}\label{eq3-18}
  \frac{3}{2}<\Re\, s<\frac{3}{2}+1\,.
\end{equation}
Cheeger has proposed to overcome this difficulty in the following
way \cite{Ch}. The analytic continuation should be constructed
separately for the first term with $n=0$ in Eq. (\ref{eq3-14}) and
for the rest sum in this equation. Further the obtained results
should be added. Obviously, such a continuation is unique.

Following the paper \cite{ZCV}, we introduce the function
\begin{equation}\label{eq3-19}
  G_{2\alpha}(s)=\sum_{n=1}^\infty
  \frac{\Gamma(np-s+1)}{\Gamma(np+s)}\,,
  \quad  \Re \, s > 1\,.
\end{equation}
For $\alpha=\pi$ ($p=1$) we have
 \begin{eqnarray}
 G_{2\pi}(s)&=&\sum_{n=1}^{\infty}\frac{\Gamma(n-s+1)}{\Gamma(n+s)}
 =\frac{\Gamma(1-s)}{\Gamma(s)}\left[1+\frac{1-s}{s}+
\frac{(1-s)(2-s)}{s(1+s)}+
 \ldots -1\right] \nonumber \\
 &=& \frac{\Gamma(1-s)}{\Gamma(s)}\left[ F(1-s, 1; s ; 1)-1\right ],
\quad \Re \, s>1, \nonumber
 \end{eqnarray}
where $F(\alpha, \beta; \gamma; z)$ is the hypergeometric
function. Equation 9.122.1 from Ref.\ \cite{GR} gives
 \[
 F(1-s, 1; s ;
 1)=\frac{\Gamma(s)\Gamma(2s-2)}{\Gamma(2s-1)\Gamma(s-1)}\,, \quad
 \Re \, s>1\,.
 \]
Finally we obtain
\begin{equation}\label{eq3-20}
  G_{2\pi}(s)=-\frac{\Gamma(1-s)}{2\Gamma(s)}\,,
  \quad \Re\, s>1\,.
\end{equation}
The right-hand side of this equation can naturally be considered
as an analytic continuation of the function $G_{2\pi}(s)$ to all
over the complex plane $s$.
On the other hand, the expression in the
right-hand side of Eq. (\ref{eq3-20}) for $1/2<\Re \, s<1$ is
nothing else as the first term with $n=0$ in the sum in Eq.
(\ref{eq3-14}). Thus the analytical continuation of the first term
in the sum (\ref{eq3-14}) is trivial.

The basic idea of the analytical continuation of the function
$G_{2\alpha}(s)$ for $\alpha\neq \pi$ proposed by Cheeger
\cite{Ch} is also simple. It relies on the fact that the $n$-th
term of the series (\ref{eq3-19}) has a polynomial asymptotics for
large $n$
\begin{equation}\label{eq3-21}
a_n=  \frac{\Gamma(np-s+1)}{\Gamma(np+s)}\sim
  (np)^{1-2s}\sum_{j=0}^\infty c_j(s)(np)^{-2j}\,,
\end{equation}
where the functions $c_j(s)$ are calculated exactly upon
substituting in Eq. (\ref{eq3-21}) the know asymptotics for the
gamma function \cite{ZCV,Olver}. From Eq.\ (\ref{eq3-21}) it follows
immediately that the series in Eq. (\ref{eq3-19}) determining the
function $G_{2\alpha}(s)$ is convergent for
\begin{equation}\label{eq3-22}
  \Re\, s> 1\,.
\end{equation}
When summing over $n$, the terms in the asymptotics (\ref{eq3-21})
with
\begin{equation}\label{eq3-23}
  \Re\, (2s+2j-1)\leq 1
\end{equation}
lead to the divergences. Therefore, it is natural to substitute
these sums by respective values of the Riemann zeta function
$\zeta_R(s)$ \cite{ZCV}
\begin{equation}\label{eq3-24}
  G_{2\alpha}(s)=p^{1-2s}\sum_{0\leq j\leq 1-\mbox{{\scriptsize Re\,}}
  s}p^{-2j}c_j(s)\, \zeta_R(2s+2j-1)+\mbox{analytical part}\,,
\end{equation}
where the analytical part is the sum of the differences between
$a_n$ and the asymptotic terms in Eq. (\ref{eq3-18}) picked out in
the way described above. In the general case, it is impossible to
calculate this analytical part, i.e., to express it in terms of
known analytical functions. However, for
\begin{equation}\label{eq3-25}
  s=\frac{1}{2}, 0, -\frac{1}{2}, -1, -\frac{3}{2}, \ldots, -\frac{k}{2}
\end{equation}
this part vanishes because for these values of $s$ the ratio
$a_n$ in Eq.\ (\ref{eq3-21}) reduces to the finite polynomial in
$n$. The positive integer $k$ in Eq. (\ref{eq3-25}) obviously
depends on the number of asymptotic terms in Eq. (\ref{eq3-25})
summed by the Riemann zeta function in the course of analytical
continuation. Thus for values of the variable $s$ listed in Eq.
(\ref{eq3-19}) there is an explicit representation of the function
$G_{2\alpha}(s)$ in terms of the Riemann zeta function. For the
boundary value $s=1$ the series defining the function
$G_{2\alpha}(s)$ diverges
\begin{equation}\label{eq3-26}
  G_{2\alpha}(1)=\sum_{n=1}^\infty \frac{\Gamma(np)}{\Gamma(np+1)}=
  \frac{1}{p}\sum_{n=1}^\infty \frac{1}{n}\,.
\end{equation}
It leads to a simple pole at the point $s=1$ which acquires the
analytic continuation of the function $G_{2\alpha}(s)$ to the
region $\Re\, s<1$.

In view of Eqs. (\ref{eq3-19}) and (\ref{eq3-20}), the zeta function
(\ref{eq3-14}) can be represented in the form
\begin{equation}\label{eq3-27}
  \zeta(s, r)=\frac{1}{2\pi \alpha
  r^{3-2s}}\frac{\Gamma(s-1)}{\Gamma(s)}\left[G_{2\alpha}(s-1/2)
-G_{2\pi}(s-1/2)\right]=\frac{1}{2\sqrt \pi \alpha r^{3-2s}}
\frac{I_{2\alpha}(s-1/2)}{\Gamma(s)}{,}
\end{equation}
where, following Ref.\ \cite{ZCV}, we have introduced the function
\begin{equation}
I_{2\alpha}(s)=\frac{\Gamma(s-1/2)}{\sqrt \pi}\left [
G_{2\alpha }(s) -G_{2\pi}(s)
\right ]{.}
\label{eq3-27a}
\end{equation}
This function, as well as $G_{2\alpha (s)}$, possesses only a simple
pole at the point $s=1$. At the first sight the gamma function
$\Gamma (s-1/2)$, entering in the definition
of  $I_{2\alpha}(s)$,    leads to the poles for $s=1/2,-1/2, -3/2, 
\ldots$~.  However, it can be shown \cite{ZCV} that the difference 
$G_{2\alpha}(s)- G_{2\pi}(s)$ vanishes at these points.

The local zeta function (\ref{eq3-27}) for electromagnetic field
inside a perfectly conducting dihedral of opening angle $\alpha $
is equal to twice the corresponding zeta function for a
massless scalar field on the space $C_\beta \times {\Bbb R}^1$ with 
$\beta =2 \alpha$, where $C_\beta $ is a cone of angle $\beta$ 
\cite{ZCV}.  This conclusion is obvious if one takes into account the 
relation between corresponding boundary value problems (see Eq.\ (12) 
in Ref.\ \cite{Dowker}). It is also easy to arrive at this conclusion 
directly writing out a complete set of the eigenfunction for a scalar 
massless field on the manifold $C_\beta\times {\Bbb R}^1$:
\begin{equation}
\label{}
f_{\lambda n k}(r,\theta, z)= \eta _{n0} \frac{e^{ikz}}{\sqrt{\pi \beta}}
 J_{np}(\lambda r){\sin \atop \cos}
(np\theta),\quad n=0,1,2,\ldots,\quad p=\frac{2\pi}{\beta}{.}
\end{equation}
In this case all the eigenvalues with $n=1,2,\ldots$ turn out to be 
double degenerate, each of them has two eigenfunctions proportional 
to $\sin (np\theta )$ or $\cos (np\theta)$. The local zeta functions 
for scalar fields inside a wedge with Dirichlet and Neumann boundary 
conditions are constructed in The Appendix.

In order to find in the approach developed the spatial distribution of the
vacuum energy inside the conducting wedge (see Eqs. (\ref{eq3-7})
and (\ref{eq3-17})) the value of $G_{2\alpha}(-1)$ should be
calculated. Here we show how to do this applying the procedure of
analytic continuation of the function $G_{2\alpha}(s)$ described
above.

{} From Eq. (\ref{eq3-23}) it follows that for $s=-1$ three terms
with $j=0, 1, 2$ in the asymptotics (\ref{eq3-21}) should be summed
by making use of the Riemann zeta function. For $s=-1$ Eq.
(\ref{eq3-21})  gives
\begin{equation}\label{eq3-28}
  \frac{\Gamma(np+2)}{\Gamma(np-1)}=
  (np)^3\left[1-\frac{1}{(np)^2}\right]\,.
\end{equation}
Comparing Eqs. (\ref{eq3-28}) and (\ref{eq3-21}) we infer
\begin{equation}\label{eq3-29}
  c_0(-1)=1\,,\quad c_1(-1)=-1\,, \quad c_2(-1)=0\,.
\end{equation}
According to Eq. (\ref{eq3-24}) the analytic continuation of the
function $G_{2\alpha}(s)$ at the point $s=-1$ has the value
\begin{eqnarray}
 G_{2\alpha}(-1)&=& \lim_{s\to
 -1}p^{1-2s}\sum_{j=0}^{2}c_j(s)p^{-2j}\zeta_R(2j+2s-1) \nonumber
 \\
 &=&
 p^3\left[c_0(-1)\zeta_R(-3)+c_1(-1)p^{-2}\zeta_R(-1)+p^{-4}\lim_{s\to
 -1}c_2(s)\zeta_R(2s+3)\right]\,.  \label{eq3-30}
\end{eqnarray}
In order to find the limit in the last term in Eq. (\ref{eq3-30}),
the exact form of the polynomial $c_2(s)$ should be used \cite{ZCV}
 \[
 c_2(s)=\frac{s}{18}\left(s^2-\frac{1}{4}\right)(s^2-1)\left(
s-\frac{6}{5}\right){,}
 \]
as well as the behaviour of the Riemann zeta function near the pole
\cite{GR}
 \[
 \zeta_R(2s+3)\simeq \frac{1}{2s+2}+\gamma+\ldots\,, \quad s\to -1\,{,}
 \]
where $\gamma$ is the Euler constant. It gives
\begin{equation}\label{eq3-31}
  p^{-4}\lim_{s\to -1} c_2(s)\zeta_R(2s+3)=-\frac{11}{120}\,
  p^{-4}\,.
\end{equation}
Substituting Eq. (\ref{eq3-31}) and the values of the Riemann zeta
function
 \[
 \zeta_R(-1)=-\frac{1}{12}\,, \quad \zeta_R(-3)=\frac{1}{120}
 \]
into Eq. (\ref{eq3-30}) one obtains
\begin{equation}\label{eq3-32}
  G_{2\alpha}(-1)=\frac{1}{120}\left(p^3+10p-\frac{11}{p}\right)=
  \frac{1}{120p}\, (p^2-1)(p^2+11)\,.
\end{equation}
Equation (\ref{eq3-20}) gives for $G_{2\pi}(-1)$
\begin{equation}\label{eq3-33}
 G_{2\pi}(-1)=0\,.
\end{equation}
Let us remind that upon analytic continuation of the function
$G_{2\pi}(s)$ the right-hand side of Eq. (\ref{eq3-20}) is
considered all over the plane of complex variables $s$.
Substituting Eqs. (\ref{eq3-32}) and (\ref{eq3-33}) into Eq.
(\ref{eq3-27}) and taking into account that
 \[
 \frac{\Gamma(-3/2)}{\Gamma(-1/2)}=-\frac{2}{3}\,,
 \]
we find
\begin{equation}\label{eq3-34}
  \zeta\left(-\frac{1}{2}, r\right)=-\frac{1}{360\pi^2r^4}\,
  (p^2-1)(p^2+11)
\end{equation}
and respectively for the density of the vacuum energy
\begin{equation}\label{eq3-35}
  \rho(r)=\frac{1}{2}\,\zeta\left(-\frac{1}{2}, r\right)=
  -\frac{1}{720\pi^2r^4}\,(p^2-1)(p^2+11)\,, \quad
  p=\frac{\pi}{\alpha}\,.
\end{equation}
For $\alpha<\pi$ ($p>1$) the vacuum energy density is negative,
and it vanishes for $\alpha=\pi$ ($p=1$), when the wedge
transforms into semispace. Putting in Eq. (\ref{eq3-35}) $\alpha\to
0$ and $r\to \infty$ in such a way that $r\alpha=d=$ constant, one
arrives at the vacuum energy density between two parallel
perfectly conducting plates
\begin{equation}\label{eq3-36}
  \rho_{plates}=-\frac{\pi^2}{720\,d^4}\,.
\end{equation}
Multiplying this expression by the volume of space bounded by
plates $S\,d$, where $S$ is the area of the plate surface, we obtain
the well known Casimir energy \cite{MTr,PM}
\begin{equation}\label{eq3-37}
  E_C=-\frac{\pi^2 S}{720 \, d^3}\,.
\end{equation}

The vacuum energy density (\ref{eq3-35}) posses  non-integrable
singularity at the point $r=0$. Here we shall not touch the
physical origin of this singularity. It is discussed in the
literature for a long time (see, for example, Refs.\ \cite{DC,BL,KCD}).
When defining the global quantities in the problem at hand,
we simply introduce a cutoff at the lower limit of integration
over $r$. Thus for the total energy we have
\begin{equation}\label{eq3-38}
  E=\int_{\varepsilon}^{\infty}rdr\int_0^\alpha d\theta \, \rho(r)=-
  \alpha\,\frac{(p^2-1)(p^2+11)}{1440\pi^2\varepsilon^2}\,.
\end{equation}

Having calculated the energy density for a conducting wedge, one
can immediately recover all the components of the relevant
energy-momentum tensor $T_{\mu\nu}$ \cite{DC}. In fact, taking
into account the symmetries of the problem and the vanishing of
the trace and divergence of
$T_{\mu\nu}$ one finds
\begin{equation}\label{eq3-39}
  T_{\mu\nu}=\frac{1}{r^4}\, f(\theta, \alpha)\,\mbox{diag}(1, -3,
  1, 1)=(T_{rr}, T_{\theta\theta}, T_{zz}, -\rho)
\end{equation}
with some function $f(\theta, \alpha)$, vanishing
when $\alpha=\pi$. Substituting
$\rho$ on the right-hand side of Eq. (\ref{eq3-39}) by Eq. (\ref{eq3-35})
we obtain
\[
  f(\alpha)=\frac{(p^2-1)(p^2+11)}{720 \pi^2}\,.
\]
Thus all the components of the energy-momentum tensor are found.
Specifically, we conclude that the surface density of the Casimir
force $F(r)$ acting on the wedge faces is
\begin{equation}\label{eq3-40}
  F(r)=-T_{\theta\theta}(r)=\frac{(p^2-1)(p^2+11)}{240\pi^2r^4}\,.
\end{equation}
These forces tend to diminish the opening angle $\alpha$ between the
wedge plates.

The torque of the Casimir forces about the origin is
\begin{equation}\label{eq3-41}
  M=\int_\varepsilon^\infty F(r)\,r\, dr=\frac{(p^2-1)(p^2+11)}{480
  \pi^2\varepsilon^2}=-\frac{3E}{\alpha}\,.
\end{equation}
Certainly it depends on the cutoff parameter $\varepsilon$. We
again recall that all the quantities found in our study are related
to a unit length along the $z$ direction.

\section{High Temperature Asymptotics of the Casimir Effect for a
Conducting Wedge}

In this section we  briefly consider the application of the constructed
zeta function to investigation of the temperature behavior of
the problem under consideration. In the general case, to find the
temperature dependence of the local characteristics proves to be
a rather complicated problem that still waits its complete solution
\cite{KCD}. Therefore we address the global thermodynamical
function of electromagnetic filed inside a perfectly conducting wedge
(the Helmhotz free energy).
For this purpose we need the global
zeta function. However Eq. (\ref{eq3-6}) cannot be used for this
purpose directly due to non-integrable singularity at $r=0$ of the
local zeta function (\ref{eq3-27}). We again introduce a cutoff in
the $r$ integral
\begin{eqnarray}
  \zeta(s)&=&\int_\varepsilon^\infty rdr\int_0^\alpha  d\theta \,
  \zeta(s, r) \nonumber \\
  &=&\frac{1}{2\pi \varepsilon^{1-2s}(1-2s)}\,
  \frac{\Gamma(s-1)}{\Gamma(s)}\left[G_{2\alpha}(s-1/2)-G_{2\pi}(s-1/2)\right]
\nonumber \\
&=& \frac{I_{2\alpha}(s-1/2)}{2\sqrt \pi \varepsilon ^{1-2s}(1-2s)\Gamma(s)}
\,.
 \label{eq4-1}
\end{eqnarray}

The appearance of the cutoff here can be formally explained in
the following way. In the general case, the local zeta function
(\ref{eq3-5}) is, from the mathematical stand point, a distribution.
Therefore its complete definition requires specification of
the relevant set of test functions. In the problem at hand
one can choose the test functions proportional to the step
function $\theta(r-\varepsilon)$. Certainly this consideration is
a formal one, and the problem of constructing the global zeta
function for boundary conditions in question  still waits its solution.

One could try to substitute the known spectrum (\ref{eq2-4})
directly in the definition of the global
zeta function (\ref{eq3-6}). However, without introducing a cutoff the
integration cannot be done here
Really,
\begin{equation}
\label{eq4-1a}
\bar \zeta (s) =\int^{\infty}_{-\infty}\frac{dk}{2\pi}\int_{0}^{\infty}
\frac{\lambda\,d\lambda}{(k^2+\lambda^2)^s}=\frac{1}{2\sqrt{\pi}}
\frac{\Gamma(s-1/2)}{\Gamma(s)}\int_{0}^{\infty}\lambda^{2-2s}d\lambda\,{.}
\end{equation}
On introducing the exponential regularization in the $\lambda$-integral we
obtain
\begin{equation}
\label{eq4-1b}
\int^\infty_0\lambda^{2-2s}d\lambda\to \int^\infty_0 e^{-\varepsilon \lambda}
\lambda^{2-2s}d\lambda=\frac{\Gamma(3-2s)}{\varepsilon^{3-2s}}{.}
\end{equation}
Finally the zeta function $\bar \zeta(s)$ acquires the form
\begin{equation}
\label{eq4-1c}
\bar \zeta (s)=\frac{1}{2\sqrt \pi}\frac{\Gamma(s-1/2)}{\Gamma(s)}
\frac{\Gamma(3-2s)}{\varepsilon ^{3-2s}}{.}
\end{equation}
This expression disagrees with Eq.\ (\ref{eq4-1}). The substantial 
shortcoming of the formula (\ref{eq4-1c}), in comparison with Eq.\ 
(\ref{eq4-1}), is independence of the opening angle $\alpha$. 
Therefore we shall use further the zeta function (\ref{eq4-1}).

     The global zeta function (\ref{eq4-1}) enables one to construct
immediately the high temperature asymptotics of the thermodynamical
functions in the problem at hand~\cite{BNP}. Here it is worth
reminding that in our consideration we are dealing only with the
spatial part of the differential operator governing the fluctuation
dynamics of electromagnetic field (see Eqs.\ (\ref{eq2-2}) and
(\ref{eq2-3})).  Therefore the corresponding global zeta function
enables one to construct, in a direct way, only the high temperature
asymptotics of the thermodynamical  characteristics in the problem
under study. In order to investigate complete temperature behavior
of these characteristics a complete spectral zeta function is needed
that involves also the summation over the Matsubara frequencies.

The high temperature expansion for the Helmholtz free energy has the
form \cite{DK} $(\hbar=c=1)$
\begin{equation}
\label{eq4-2}
F(T)\simeq \frac{T}{2}\zeta '(0)+\sum_{n=0,1/2,1,3/2, \ldots}
a_nT^{4-2n}c_n, \quad T\to \infty\,{.}
\end{equation}
Here $c_n$'s are numerical coefficients
\begin{equation}
\label{eq4-3}
c_0=-\frac{\pi^2}{90},\quad c_{1/2}=-\frac{\zeta_R(3)}{4\pi^{3/2}}, 
\quad c_1=
-\frac{1}{24}, \quad \cdots ,
\end{equation}
safe for $c_{3/2}$   and $c_2$, which also depend on $\ln T$, and $a_n$'s are
the heat kernel coefficients. The latter are determined  by the zeta
function through the relation
\begin{equation}
\label{eq4-5}
\frac{a_n}{(4\pi)^{3/2}}=\lim_{s\to\frac{3}{2}-n}\left (
s+n-\frac{3}{2}
\right )\zeta (s)\Gamma (s), \quad n=0, 1/2,1, \ldots \,{.}
\end{equation}

     As follows from Eq.\ (\ref{eq4-1}) the zeta function $\zeta (s)$ 
has simple poles at the point $s=1/2$ (due to the multiplier $(1-2s)$ 
in the denominator) and at the point $s=3/2$ because the function 
$I_{2\alpha}(z)$ has such a pole at the point $z=1$. Thus, only the 
heat kernel coefficients $a_0$ and $a_1$ do not vanish in the problem 
under study.  This is in  full accord with the general properties of 
the heat kernel expansion \cite{Bordag}. Indeed, for flat manifolds 
without boundary or with flat boundary all the heat kernel 
coefficients except for $a_0$ are equal to zero.

And what is more, one can show in the general case that the heat
kernel coefficient $a_0$ is equal to the volume of the system under
consideration. The corresponding term in the high temperature
expansion (\ref{eq4-2})
\[
-a_0\frac{\pi^2}{90}T^4
\]
is the Stefan-Boltzmann law for the free energy of electromagnetic
field in the space restricted  by the boundaries
\begin{equation}
\label{eq4-6}
\frac{F}{V}=-\frac{\pi^2}{90}T^4=-\frac{2}{3}\sigma T^4\,{,}
\end{equation}
where $\sigma= \pi^2 k^4_B/(60 c^2\hbar ^3)$ is the Stefan-Boltzmann constant.

When the system has an infinite volume (just that very case is 
considered here), a
special limiting procedure should be used in calculations of $a_0$
(see, for example, Ref.\ \cite{SW}). In the problem at hand this
implies  the introduction of  a cutoff also at large values of $r$:
$\varepsilon \le r\le R$. In view of this the substitution of the
zeta function (\ref{eq4-1}) into Eq.\ (\ref{eq4-5}) gives
\begin{equation}
\label{eq4-7}
a_0= (R^2-\varepsilon ^2)(\alpha - \pi)\,{.}
\end{equation}
In the problem at hand we are dealing with two independent scalar
fields. Therefore the value (\ref{eq4-7}) should be compared with
doubled volume of the system under study
\begin{equation}
\label{eq4-8}
2V=2\pi
(R^2-\varepsilon^2)\frac{\alpha}{2\pi}=(R^2-\varepsilon^2)\,\alpha\,{.}
\end{equation}
It is worth noting once more that all our formulas give the 
quantities per unit length along the axes OZ. The distinction between 
Eqs.\ (\ref{eq4-7}) and (\ref{eq4-8}) are obviously attributed to a 
special condition satisfied by the zeta function (\ref{eq4-1}), 
namely, $\zeta (s)=0$ for $\alpha=\pi$.

All the terms in the asymptotics (\ref{eq4-2}), safe for the term 
with $n=0$, describe the deviation from the Stefan-Boltzmann law. In 
the problem in question the coefficient $a_1$ does not vanish due to 
the discontinuity of the boundary at the point $r=0$ (wedge or cone 
singularity). There is a voluminous  literature concerning with such 
a singularity. we mention here only a few papers 
\cite{ZCV,Ch,Wedge1,Wedge2,Fur1,Fur2} where further references can be 
found.

Taking into account the behaviour of the  gamma function at the origin
$\Gamma (s)\sim s^{-1}, \quad s\to 0$ we deduce from Eq.\ (\ref{eq4-1})
\begin{equation}
\label{eq4-9}
\zeta ' (0)=\frac{I_{2\alpha}(-1/2)}{2\sqrt \pi \varepsilon}{.}
\end{equation}
The explicit formulas derived in Ref. \cite{ZCV} give
\begin{equation}
\label{eq4-10}
I_{2\alpha}(-1/2)=-\frac{\zeta '(-2)}{2\sqrt \pi}
=\frac{\zeta_R(3)}{8\pi^2\sqrt \pi}{.}
\end{equation}
Hence
\begin{equation}
\label{eq4-11}
\zeta'(0)=\frac{\zeta_R(3)}{16\pi ^3\varepsilon}{.}
\end{equation}

Substituting the zeta function (\ref{eq4-1}) in Eq.\ (\ref{eq4-5}) we find
\begin{equation}
\label{eq4-12}
\frac{a_1}{(4\pi)^{3/2}} =\lim_{s\to 1/2}(s-1/2)\zeta(s)\Gamma(s) =
-\frac{I_{2\alpha}(0)}{4\sqrt \pi}{.}
\end{equation}
In order to calculate $I_{2\alpha}(0)$ we take, in accord with 
relation (\ref{eq3-23}), two
terms in Eq.\ (\ref{eq3-24}) with $j=0,1$. It gives
\begin{equation}
\label{eq4-13}
G_{2\alpha}(0)=\lim_{s\to 0}p^{1-2s}[c_0(s)\zeta_R(2s-1)-p^{-2}
c_1(s)\zeta_R(2s+1)]{.}
\end{equation}
The coefficients $c_0(s)$  and $c_1(s)$ are explicitly written, for example,
in Ref.\ \cite{Olver}
\begin{equation}
\label{eq4-14}
c_0(s)=1, \quad c_1(s)=\frac{s}{3}(s-1/2)(s-1)\,{.}
\end{equation}
With allowance for the behaviour of the Riemann zeta function close by pole
\[
\zeta_R(1+2s)\simeq \frac{1}{2s}+\gamma+\cdots, \quad s\to 0
\]
we obtain
\begin{equation}
\label{eq4-15}
G_{2\alpha}(0)=\frac{1}{12}(p^{-1}-p)\,{.}
\end{equation}
Form Eq.\ (\ref{eq3-20}) it follows that
\begin{equation}
\label{eq4-16}
G_{2\pi}(0)=0\,{.}
\end{equation}
Finally we have
\[
I_{2\alpha}(0)=\frac{1}{6}(p-p^{-1})=\frac{1}{6}\left (
\frac{\pi}{\alpha}-\frac{\alpha}{\pi}
\right )
\]
and
\begin{equation}
\label{eq4-17}
a_1=-2\pi I_{2\alpha }(0)=-\frac{\pi}{3}\left (
\frac{\pi }{\alpha} -\frac{\alpha}{\pi}
\right ){.}
\end{equation}

Now we are in position to write the high temperature corrections
to the Stefan-Boltzmann law inside a conducting dihedral angle
\begin{equation}
\label{eq4-19}
F_{cor}(T)\simeq-\frac{T\zeta_R(3)}{32
\pi^3\varepsilon}+\frac{\pi^2-\alpha^2}{72 \alpha\hbar c} T^2, \quad
T\to \infty \,{.}
\end{equation}
Here the constants $\hbar $ and $c$ are explicitly restored in order
to separate a pure classical contribution (the first term on the
right hand side of Eq.\ (\ref{eq4-19})) and the quantum correction
(the second term in Eq.\ (\ref{eq4-19})).

   The authors of  Ref.\ \cite{BD} argue that the high
temperature asymptotics of the free energy of electromagnetic field
inside a perfectly conducting wedge should be the same as that for
smooth conducting surfaces with nonzero curvature
\begin{equation}
\label{eq4-20}
F(T)\simeq CT\ln (T/Q), \quad T\to \infty,
\end{equation}
with $C$ and $Q$ being some constants\footnote{The examples of such
asymptotics can be found in Ref.\ \cite{BNP}}. The specific term
proportional to $T^2$ in the asymptotics (\ref{eq4-20}) was not
brought out.

Making use of Eq.\ (\ref{eq4-19}) one can try to  estimate, at least
in kind, the behaviour of the torque of the Casimir forces at high
temperature. For this purpose we substitute the total energy $E$ in
Eq.\ (\ref{eq3-41}) by the asymptotic expression for the free energy
in the problem under study
\begin{equation}
\label{eq4-21}
M\simeq \frac{3\zeta_R(3)T}{32\pi ^3 \alpha\varepsilon}-
\frac{\pi^2-\alpha^2}{24 \alpha^2}\frac{T^2}{\hbar c}, \quad T\to
\infty \,{.}
\end{equation}
The book \cite{LR} is a sole source where the high temperature
behaviour of the torque $M$ has been considered. The authors have
used the method of images for constructing the Green's functions of
electromagnetic field inside the dihedral, the photon energy being
cut in ultraviolet region. In terms of the notations used in the
present paper, the high temperature asymptotics of $M$ found in Ref.\
\cite{LR} reads
\begin{equation}
\label{eq4-22}
M\simeq \frac{\pi T}{12 \alpha^2\varepsilon}, \quad T\to \infty.
\end{equation}
Here we have also to ascertain disagreement with our calculations.

In the preceding section it was shown that the zeta function of
electromagnetic field  inside a perfectly conducting dihedral of
opening angle $\alpha $  is equal to double zeta function for a
massless scalar field on the manifold $C_\beta\times {\Bbb R}^1$ with
$2 \alpha=\beta$, where $C_\beta$ is a cone of angle $\beta$.
Therefore all the results obtained in our consideration are directly
applicable to the massless scalar field on the manifold
$C_\beta\times {\Bbb R}^1$, in particular to the cosmic string
background.  Really, the general
solution to the Maxwell equations in the presence of a cosmic string
are expressed in terms of two scalar massless fields that satisfy the
periodicity conditions in angle variable with a period $\beta =2\pi
-\phi $, where  $\phi $ is the deficiency angle of the metric due to
the cosmic string, $\phi =8 \pi G \mu $.  Most simply one can come to
this conclusion in the following way. Let us assume that the string
has a finite transverse size, $r_{str}$, and its surface is perfectly
conducting.  Obviously, the  reasonings of the Section II apply to
this boundary configuration, two scalar functions $U$ and $V$ being
periodic in angle variable $\theta $ and meeting the Dirichlet and
Neumann boundary conditions at $r=r_{str}$. Letting $r_{str}$ tend to zero,
we substitute the Dirichlet and Neumann boundary conditions at $r=r_{str}$ by
requirement that the relevant solutions to be bounded everywhere including
the origin: $0\leq r< \infty$. The periodicity condition in angle variable
$\theta$ apparently remains. From here we infer that the local zeta 
function for electromagnetic field in the presence of cosmic string 
is obtained from the relevant zeta function for the conducting wedge 
(\ref{eq3-27}) by the substitution $2\alpha = 2\pi - 8\pi G\mu$. 
Specifically, for the vacuum energy density in this case we have 
again Eq.\ (\ref{eq3-36}) with the same substitution.  The same 
result was obtained in Ref.\ \cite{FS} by making use of the Greens 
function method.  Further, the high temperature corrections to the 
Stefan-Boltzmann law for the free energy of electromagnetic field on 
the background of a cosmic string  are given by the formula
\begin{equation}
\label{eq4-23}
F(T)\simeq \frac{T\zeta_R(3)}{32 \pi^3 \varepsilon}+\frac{T^2}{72
\hbar c} \frac{\phi}{2}\frac{4\pi -\phi}{2\pi -\phi}, \quad T\to
\infty {.}
 \end{equation}

\section{Conclusion}
We have shown that the technique of the local zeta function enables 
one to carry out, in a consistent way, the calculation of the vacuum 
energy density of electromagnetic field inside a conducting wedge 
without dealing with obvious divergences. This regularization method 
leads to a finite expression for the energy density under study 
without any subtractions. Our consideration is a natural addition to 
studies of the Casimir effect for a wedge accomplished in recent 
papers \cite{BL,BP} by making use of Green's function method.  
Employment of the Hertz potentials for constructing the general 
solution of the Maxwell equations results in a considerable 
simplification of the calculations of the local zeta function in the 
problem at hand. Transition to the global zeta function has been 
carried out by introducing a cutoff for small $r$. On this basis the 
nonzero heat kernel coefficients $a_0$, $a_1$, and the zeta 
determinant were calculated. Remarkably that the coefficient $a_1$ 
does not depend on the cutoff and it is completely determined by the 
wedge singularity of the boundary. Proceeding from this the high 
temperature asymptotics of the Helmholtz free energy and of the 
torque of the Casimir forces are found. The wedge singularity gives 
rise to a specific high temperature behaviour $\sim T^2$ of the 
quantities under consideration. The found results are directly 
applicable to the free energy of electromagnetic field on the 
background of a cosmic string.

\acknowledgments
V.V.N.\ is grateful to  Dr.~D.~V.~Fursaev for elucidating discussion of the
cone singularity and its physical implication and to
Dr.~I.~G.~Pirozhenko for reading manuscript and correcting errors
in formulas.  This work was
accomplished during the visit of V.V.N. to Salerno University. It is
a pleasure for him to thank Professor G.\ Scarpetta, Drs.  G.\
Lambiase and A. \ Feoli for warm hospitality.  V.V.N.\ acknowledges
the financial support of INFN, Russian Foundation for Basic Research
(Grant No.\ 00-01-00300), and International Science and Technology
Center (Project No.\ 840).
\appendix
\section*{Local zeta functions for scalar massless fields inside
dihedral angle with Dirichlet and
Neumann boundary conditions}
First we consider the Dirichlet boundary conditions with the
eigenfunctions (\ref{eq2-5}).  The local zeta function
is given by
\begin{eqnarray}
\zeta_D (s,r,\theta)&=& \frac{1}{\pi \alpha}\int_{-\infty}^\infty dk
\int_{0}^{\infty}\frac{\lambda\,d\lambda}{(k^2+s^2)^s}\sum_{n=1}^{\infty}
J_{np}^2(\lambda r)\sin ^2(np\theta) \nonumber \\
&=& \frac{\Gamma (s-1/2)}{2\sqrt \pi \alpha\Gamma (s)}
\int_{0}^{\infty}d\lambda\,
\lambda^{2-2s}\sum_{n=1}^{\infty}J_{np}^2(\lambda r)
[1-\cos (2np)]\,{.}
\label{A1}
\end{eqnarray}
By mens of the formula 8.531.3 from \cite{GR}
\[
J_0(z\sin \alpha)=J_0^2
\left (
                             \frac{z}{2}
\right )+2\sum_{k=1}^{\infty}J_k^2
\left (
                             \frac{z}{2}
\right )
\cos(2 k\alpha)
\]
the dependence on the angle $\theta $ in Eq.\ (\ref{A1}) can be 
stored in a single
term
\begin{eqnarray}
\zeta_D(s,r,\theta) &=&\frac{\Gamma (s-1/2)}{2\sqrt \pi \alpha\Gamma (s)}
\int_{0}^{\infty}d\lambda\, \lambda^{2-2s}\left [
\frac{1}{2}J_0^2(\lambda r)+\sum_{n=1}^{\infty}J^2_{np}(\lambda r)-
\frac{1}{2}J_0(2 \lambda r \sin p\theta)
\right ]
\nonumber \\
&=& \frac{1}{2}\zeta(s,r) -\frac{\Gamma (s-1/2)}{4\sqrt \pi \alpha \Gamma}
\int_{0}^{\infty}d\lambda \,\lambda^{2-2s}J_0(2\lambda r \sin(p\theta))\,{,}
\label{A2}
\end{eqnarray}
where $\zeta(s,r)$ is the local zeta function for electromagnetic field
in a conducting
wedge defined in Eq.\ (\ref{eq3-9}). By making use of the integration formula
6.561.14 from \cite{GR}
\[
\int_{0}^{\infty}x^\mu J_\nu(a x)dx = 2^\mu a^{-\mu-1}\frac{
\Gamma\left
(\frac{1}{2}+\frac{1}{2}\nu +\frac{1}{2}\mu
\right )
}{
\Gamma\left
(\frac{1}{2}+\frac{1}{2}\nu- \frac{1}{2}\mu
\right )
}
\quad - \Re \nu -1 <\Re \mu <\frac{1}{2}, \quad a>0
\]
we cast Eq.\ (\ref{A2}) into the final form
\begin{equation}
\label{A4}
\zeta_D(s,r,\theta)=\frac{1}{2}\zeta(s,r)+\frac{1}{8\sqrt \pi \alpha(
r \sin(p\theta))^{3-2s}}\frac{\Gamma(3/2-s)}{\Gamma(s)}\,{.}
\end{equation}
The last term in Eq.\ (\ref{A4}) is first defined in the strip
\begin{equation}
\label{A5}
\frac{3}{4}<\Re s<\frac{3}{2},
\end{equation}
determined by the conditions of applicability of Eq.\ (\ref{A4}).
As usual under the analytic
continuation, we assume that outside this strip the same equation holds.

Proceeding along similar lines we obtain the local
zeta function for the Neumann boundary conditions
\begin{equation}
\label{A6}
\zeta_D(s,r,\theta)=\frac{1}{2}\zeta(s,r)-\frac{1}{8\sqrt \pi \alpha(
r \sin(p\theta))^{3-2s}}\frac{\Gamma(3/2-s)}{\Gamma(s)}\,{.}
\end{equation}
In particular the vacuum energy densities for massless scalar fields subjected
to the Dirichlet and Neumann boundary conditions on the wage sites are
\begin{equation}
\label{A7}
\rho _D(r,\theta)=\rho_-(r,\theta), \quad
\rho _N(r,\theta)=\rho_+(r,\theta),
\end{equation}
where
\begin{equation}
\label{A8}
\rho_{\pm}(r,\theta)=\frac{1}{2}\rho(r)\pm
\frac{1}{16 \pi \alpha r^4\sin^4(p\theta)}\,{,}
\end{equation}
and $\rho(r)$ is the vacuum energy density of electromagnetic 
field, defined in Eq.\
(\ref{eq3-35}).

Thus the vacuum energy densities derived have nonintegrable 
singularities near by the wedge sites. It is a typical behaviour of 
spatial distribution of the Casimir energy (see, for example, Ref.\ 
\cite{OR}).

     The local zeta function technique, consider in the present 
paper, cannot directly be applied to the scalar field with conformal 
coupling.  The point is that in this case the relation (\ref{eq3-8a}) 
does not hold.

\end{document}